\documentclass{aa}  
\newcommand{\noise}{{\mathcal N}}

\usepackage{graphicx}
\usepackage{xcolor}
\usepackage{txfonts}
\usepackage{natbib}
\bibpunct{(}{)}{;}{a}{}{,} 
\usepackage{amsmath}
\usepackage{hyperref}
\hypersetup{
    colorlinks = true,
    linkbordercolor = {white},
    citecolor={blue},
    linkcolor ={blue},
}
\begin{document}

   \title{
  Detection of Rossby modes with even azimuthal orders using helioseismic normal-mode coupling
  }


   \author{K. Mandal,
          \inst{1}
          S. M. Hanasoge
          \inst{2,4}
        \and
          L. Gizon
          \inst{1,3,4}
         }
   \institute{Max-Planck-Institut für Sonnensystemforschung, Justus-von-Liebig-Weg 3, 37077 Göttingen, Germany\\
              \email{mandal@mps.mpg.de}\\
              \and
             Tata Institute of Fundamental Research, Mumbai 400005, India\\
             \and
             Institut für Astrophysik, Georg-August-Universität Göttingen, 37077 Göttingen, Germany\\
             \and
             Center for Space Science, New York University Abu Dhabi, PO Box 129188, Abu Dhabi, UAE
             }

   \date{Received ----; accepted ----}

 
  \abstract
   {Retrograde Rossby waves, measured to have significant amplitudes in the Sun, likely have notable implications for various solar phenomena.}
   {Rossby waves create small-amplitude, very-low frequency motions (on the order of the rotation rate and lower), which in turn shift the resonant frequencies and eigenfunctions of the acoustic modes of the Sun. The detection of even azimuthal orders
    Rossby modes using mode coupling presents additional challenges and prior work therefore only focused on odd orders. Here, we successfully extend the methodology to measure even azimuthal orders as well.}
   {We analyze $4$ and $8$ years of Helioseismic and Magnetic Imager (HMI) data  and consider coupling between different-degree acoustic modes (of separations $1$ and $3$ in harmonic degree). The technique uses couplings between different frequency bins to capture the temporal variability of the Rossby modes.}
   {We observe significant power close to the theoretical dispersion relation for sectoral Rossby modes (where the azimuthal order is same as harmonic degree, $s=\vert t \vert$). Our results are consistent with prior measurements of Rossby modes with azimuthal orders over the range $t=4$ to  $16$ with maximum power occurring at mode $t=8$. The amplitudes of these modes vary from $1$ to $2$ m/s. We place an upper bound of $0.2$ m/s on the sectoral $t=2$ mode, which we do not detect in our measurements.}
   {This effort adds credence to the mode-coupling methodology in helioseismology.}
 \titlerunning{Solar Rossby waves}
 \maketitle{}
\section{Introduction} \label{sec:intro}
 Rossby waves are named after their discoverer, Carl-Gustaf Rossby, who first explained the largest scale oscillatory motions on Earth atmosphere \citep{rossby_1939} as arising from the conservation of potential vorticity. \citet{chelton96} observed these oscillations in the ocean by analyzing variations in sea-surface height from satellite data. These large-scale motions have implications on terrestrial weather and can influence convection and differential rotation in stars if they have sufficient amplitudes \citep{plaskett66}. Rossby-like waves (also known as r-modes in the astrophysical context) can be sustained by any rotating spherical fluid body, such as the Sun, in which the Coriolis force acts as a restoring force \citep{Papa1978,provost_81,saio_82}. These waves have frequencies comparable to the rotation rate.  For a uniformly rotating star,
Rossby modes follow the dispersion relation \citep{provost_81}
 \begin{equation}
    \sigma=-\frac{2\Omega t}{s(s+1)},\label{eq:dispersion} 
 \end{equation}
 in a co-rotating frame,
 where $s$ and $t$ are the harmonic degree and azimuthal order of these modes respectively \footnote{We use positive values for azimuthal orders and negative values for frequencies to describe retrograde Rossby modes in this work. Similar convention is also considered by other authors. Note that this convention is different than the one used in our previous works \citep{mandal2020,hanasoge19}}. The availability of long-term high-resolution observational data of the Sun from Mount Wilson observatory, the Michelson Doppler Imager (MDI) on-board the Solar and Heliospheric Observatory (SOHO) and Global Oscillation Network Group (GONG) have inspired several authors \citep[e.g.][]{Kuhn2000,ulrich2001,sturrock15} to search for these oscillatory motions in the Sun. 
 \citet{sturrock15} observed oscillations in solar radius measurements and attributed them due to r modes with $t=1$. All these earlier studies lacked the measurement of the dispersion relation of these modes, critical to characterizing and  understanding them. \citet{gizon18} were the first to measure the dispersion relation of these waves in the Sun using surface-granulation tracking methods and ring-diagram analysis with $6$ years of SDO/HMI data and unambiguously detected modes with azimuthal order starting from $t=3$ to $t=15$.  The surface eigenfunctions are close to sectoral spherical harmonics, although they are more peaked about the equator as a result of the effect of differential rotation \citep{gizon2020}.
 The detection of Rossby modes was later confirmed by several authors using different methods, e.g., by \citet{liang_2018} who used time-distance helioseismology, by \citet{hanson20} and \citet{proxauf2020} who used ring-diagram analysis, and by \citet{hanasoge19} who applied normal-mode coupling \citep[for a description of the method, see][]{woodard_89,lavely92,roth2003,woodard13}.  Interested readers are referred to \citet{zaqa21} and references therein for a detailed review on Rossby waves in the Sun and other astrophysical problems. \par
 
 Although very powerful in its scope and inferential quality, the method of mode coupling is only slowly gaining traction in helioseismology \citep[see][for some other recent developments with this method]{schad2020,hanson21,samarth21}; for instance, there is limited work on characterizing its limitations and developing mitigation strategies. The detection of Rossby modes by \citet{hanasoge19} using this method added an important milestone to this methodology.  \citet{mandal2020} compared properties of these modes, e.g., frequency and line width, with prior work by \citet{gizon18,liang_2018} and developed an approach to mitigate systematical errors that arise in this method, e.g., leakage in the spatial domain because of our limited vantage of the Sun. \citet{hanasoge19,mandal2020} considered coupling between acoustic modes with same harmonic degrees. They detected Rossby modes only in the sectoral power spectra. Their measurement was not sensitive to even harmonic degrees which limited their detection to Rossby modes with odd harmonic degrees only. Here, we consider couplings between acoustic modes with different harmonic degrees and report the first detection of sectoral Rossby modes with even harmonic degrees using this method.

 \section{Data analysis} \label{sec:data}
 Line-of-sight Doppler velocity, $\Phi$, observed by space-based, e.g., SOHO/MDI \citep{scherrer95}, SDO/HMI \citep{hmi} and ground-based observatories, e.g., GONG are the main inputs to helioseismology. The data are transformed into the spherical-harmonic domain to obtain $\Phi_{\ell m}$, where $(\ell, m)$ are harmonic degree and azimuthal order of the p mode. As described earlier, we use $s$ and $t$ to indicate harmonic degree and azimuthal order of the Rossby waves to avoid confusion. A detailed discussion about how this data product is obtained from line-of-sight Doppler velocity data may be found in \citet{larson2015}. We obtain these time series from the JSOC website \footnote{JSOC: \url{http://jsoc.stanford.edu/}}. We perform a temporal Fourier transform on the data to obtain $\Phi_{\ell m}^\omega$, where $\omega$ is the temporal frequency. The next step is to perform cross correlations across wave numbers, i.e., $\Phi_{\ell m}^{\omega *}\Phi_{\ell+\Delta\ell\,\, m+t}^{\omega+\sigma}$, where $\sigma$ is varied to capture the time dependence of the perturbation and the difference between the two azimuthal order, $t$, captures the length scale of the perturbation. $\Delta\ell$ is the difference between harmonic degrees of the two acoustic modes of interest.
 For the problem of Rossby waves, we know the dispersion relation analytically for a uniformly rotating fluid body with angular rotation rate $\Omega$, which in the co-rotating frame (at the same rotational frequency), is given by Equation \ref{eq:dispersion}.
  The observed latitudinal eigenfunctions of these modes, labeled by only azimuthal order $t$, which may have contributions from sectoral $(s=\vert t\vert)$ and non-sectoral modes $(s\ne \vert t\vert)$ \citep{gizon18,proxauf2020} peak at the equator and switch sign near $\,30^{\circ}$ latitude at the surface. Latitudinal eigenfunctions of these modes can not have any zero-crossings if these modes are purely sectoral. This can either be explained by considering presence of non-sectoral modes \citep{proxauf2020} or due to the effect of differential rotation on the Rossby mode eigenfunction \citep{gizon2020}. We in this work do not attempt to answer the above  aspect of Rossby modes. Though normal mode coupling can easily distinguish between sectoral and non-sectoral Rossby modes, we in this work only focus on sectoral ones. In that case, the dispersion relation ( Equation \ref{eq:dispersion} ) simplifies to 
 \begin{equation}
    \sigma=-\frac{2\Omega}{(s+1)}.\label{eq:dispersion2} 
 \end{equation}
We choose $\Omega/2\pi=453.1$ nHz, corresponding to the equatorial rotation rate of the Sun. We analyze the first $4$ years of SDO/HMI data, from $2010$ to $2014$. We also analyze $8$ years of SDO/HMI data from $2010$ to $2018$. The reason for choosing two different data sets is to demonstrate robustness of the method, i.e., that it can capture signals from a shorter time series. Though we have access to a few decades of high-resolution data for the Sun, we cannot say the same in the context of asteroseismology. Therefore, if a method works with a shorter time series of data, it can also be easily extended to asteroseismology.  We analyze all the p-modes in the harmonic degree range $\ell\in[50,180]$ and for all identified radial orders, $n$. We slightly modify our measurements for this particular work from that of \citet{hanasoge19,mandal2020} as they considered correlations between acoustic modes with same harmonic degree which are only sensitive to odd harmonic degree of Rossby waves \citep[see Equation $6$ and $7$ of][]{hanasoge18}. For notational convenience, we use the same convention as in \citet{hanasoge18} unless otherwise mentioned. Analyzing these correlation measurements at all temporal frequencies, $\omega$, $\sigma$ and for all pairs of acoustic modes with quantum numbers $(n,\ell,m)$ is very cumbersome. To simplify the analysis, we measure the $B$-coefficient, $B^\sigma_{st}$ \citep{woodard16} which captures signal due to Rossby mode with harmonic degree, $s$ and azimuthal order, $t$ from couplings between acoustic modes with identical radial orders and $\Delta\ell=1,3$, which is defined as,
 \begin{equation}
    B^\sigma_{s\,t}(n,\ell,\ell+\Delta\ell)=\frac{\sum_{m,\,\omega}\gamma^{\ell+\,\Delta\ell\,\,s\,\ell}_{t\, m}\,H^{\sigma *}_{\ell\,\ell+\,\Delta\ell\,m\,t}(\omega)\,\Phi^{\omega *}_{\ell m}\,\Phi^{\omega+\,\sigma+\,t\Omega}_{\ell+\,\Delta\ell\,\, m+t}}{\sum_{m,\omega}\vert \gamma^{\ell+\,\Delta\ell\,\,s\,\ell}_{t\, m}\,H^{\sigma *}_{\ell\,\ell+\Delta\ell\,\,m\,t}\vert^2},\label{eq:BCoeff}
 \end{equation}
 where 
 \begin{equation}
     \gamma^{\ell+\,\Delta\ell\, s\, \ell}_{t\,m}=(-1)^{m+\,t}\sqrt{(2s+1)}\begin{pmatrix}
\ell+\Delta\ell & s & \ell\\
-(m+t) & t & m
\end{pmatrix},
 \end{equation}
 and 
\begin{multline}
 H^\sigma_{\ell\, \ell+\Delta\ell\, m\, t}(\omega) = -2\,\omega\, 
  L_{\ell\, m}^{\ell\, m}\,L_{\ell+\,\Delta\ell\,\,m+t}^{\ell+\,\Delta\ell\,\,m+t}\\\left(N_{\ell+\,\Delta\ell}\, R^{\omega*}_{\ell\, m}\, | R^{\omega+\,\sigma+\,t\Omega}_{\ell+\Delta\ell\,\, m+t}|^2 +
  N_{\ell}\,|R^{\omega}_{\ell\, m}|^2\,  R^{\omega+\,\sigma+\,t\Omega}_{\ell+\Delta\ell\,\, m+t}\right)
\end{multline}
where $N_\ell$ is the mode normalization constant, $L_{\ell\, m}^{\ell^\prime\, m^\prime}$ denotes leakage from mode $(n,\ell,m)$ to another mode $(n,\ell^\prime,m^\prime)$ due to our limited vantage of the Sun \citep{schou94} and the expression of $R^\omega_{\ell\, m}$ 
 is \citep[see Equation $11$ of][]{hanasoge17_etal},
 \begin{equation}
     R^\omega_{\ell m}=\frac{1}{(\omega_{n \ell m}-i\Gamma_{n\ell}/2)^2-\omega^2},
 \end{equation}

 where $\omega_{n\ell m}$ and $\Gamma_{n\ell}$ are the eigenfrequency and full width at half maximum of the mode, $(n,\ell,m)$.
 We analyze p-modes in the frequency range $[1000,4600]\,\, \mu$Hz. In order to capture the temporal evolution of Rossby modes, we vary $\sigma/2\pi\in[-350,0.0]$ nHz, sufficient to capture  
 the temporal evolution of all Rossby modes because its maximum frequency is $\sigma/2\pi=-302$ nHz for $(s,t)=(2,2)$ (we consider only even harmonic degrees in this work). For consistent analysis, we need to estimate noise in the measurement properly. We follow the analysis of \citet{hanasoge18} to estimate the systematic noise in these measurements, and use Equation~\ref{eq:BCoeff} to estimate variance. Taking a sum over all frequency bins $\omega$ in Equation \ref{eq:BCoeff} for all pairs of acoustic modes considered in the mode-coupling measurement is computationally very expensive. Therefore we restrict the sum to coupling between $(n,\ell,m)$ and $(n,\,\ell+\Delta\ell,\,m+t)$ modes in the frequency intervals
\begin{align}
    & \left| \omega-\omega_{n\,\ell\, m} \right| \leq \Gamma_{n\,\ell}  \\
    & \textrm{and }
    \left|\omega- \omega_{n\,\,\ell+\,\Delta\ell\,\,m+t}+(\sigma+\,t\Omega)\right|\leq 
   \Gamma_{n\,\,\ell+\,\Delta\ell} .
    \label{eq:rangeSum}
\end{align}

 The reason to subtract $\sigma+t\Omega$ in Equation \ref{eq:rangeSum} is to restrict the frequency interval to a full width around the resonance associated with $(n,\,\ell+\Delta\ell,\,m+t)$. We then connect these measured $B$-coefficients with the perturbation in the medium, which in turn may be used for inference by performing an inversion as discussed in the next section.
 
 \subsection*{Inversion}\label{sec:inversion}
 Rossby waves are described as a toroidal flow and may be expressed as  
 \begin{equation}
    u^\sigma(r,\theta,\phi)=\sum_{s,\,t}w^\sigma_{s\,t}(r)\,\hat{\mathbf{r}}\times{\boldsymbol{\nabla}}_{1} Y_{s\,t}(\theta,\phi),\label{eq:flow}.
 \end{equation}
 
\begin{figure*}
    \centering
    \includegraphics[scale=0.45]{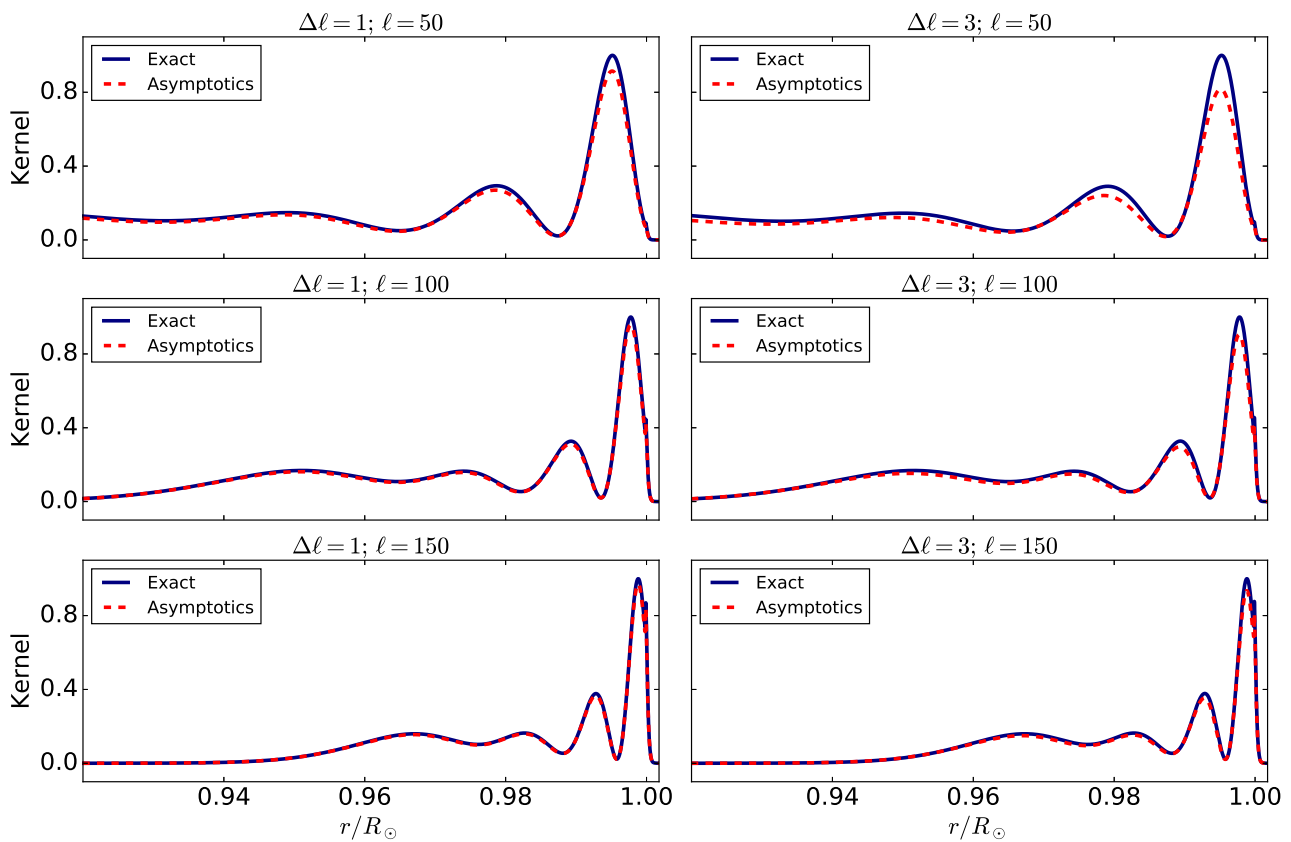}
    
    \caption{Comparison between exact \citep[equation~(53) of][]{hanasoge18} in solid navy blue and asymptotic kernels ( $\mathcal{K}_{n\ell}(r)$ multiplied by $f_{\ell^\prime-\ell,s}$) in dashed red, for $s=6$. Panels on the left and right sides are for $\ell^\prime-\ell=1$ and $3$ respectively. We consider three different harmonic degrees $\ell=50,100,150$ which cover almost the entire range of harmonic degrees that we have used for data preparation. We have normalized both kernels in each panel by the maximum values of the kernel obtained from the exact expression. Both kernels match better at larger values of $\ell$, as expected.}
    \label{fig:kernelAsymp}
\end{figure*}

 $Y_{s\,t}$ is the spherical harmonic degree with degree $s$ and azimuthal order $t$ and $\boldsymbol{\nabla}_1=\hat{\theta}\partial_\theta+\frac{\hat{\phi}}{\sin\theta}\partial_{\phi}$ where $\hat{\theta}$ and $\hat{\phi}$ are the unit vectors along increasing co-latitude, $\theta$ and longitude, $\phi$ direction. The term $w^{\sigma}_{s\,t}(r)$ captures the depth dependence of mode. We only focus on sectoral Rossby modes in this work as discussed in Section \ref{sec:data} and consider azimuthal order, $t=s$. The successful measurement of Rossby modes requires this function to attain  significant amplitudes beyond the background power in the vicinity of the theoretical dispersion relation, i.e., Equation~\ref{eq:dispersion2}. If we use the flow profile from Equation \ref{eq:flow} as a perturbation to the background model, the mode eigenfunction computed using the background model will be coupled, which, after some algebra, may be estimated as \citep[see for][]{hanasoge18}
 \begin{equation}
    B^{\sigma}_{st}(n,\ell,\ell^{\prime})=f_{\ell^{\prime}-\ell}\int_\odot dr\ \mathcal{K}_{n\ell}(r)w^{\sigma}_{st}(r), \label{eq:b_Coef}
 \end{equation}
 where $f_{\ell^\prime-\ell,s}\mathcal{K}_{n\ell}$ corresponds to the sensitivity kernel for the coupling between mode $(n,\ell,m)$ and $(n,\ell^\prime,m+t)$. $f_{\ell^\prime-\ell,s}$ is obtained from the asymptotic expression of the kernel \citep{vorontsov11} as the following 
 \begin{equation}
  f_{\ell^{\prime}-\ell}=(-1)^{(s+\ell^\prime-\ell-1)/2}\times\frac{(s-\ell^\prime+\ell)!!(s+\ell^\prime-\ell)!!}{\sqrt{(s-\ell^\prime+\ell)!(s+\ell^\prime-\ell)!}}
  \label{eq:fS}
 \end{equation}
 for odd $s+\ell^\prime-\ell$.
  This form of the kernel is valid when $s\ll \ell$ or $s\ll \ell^\prime$. The usefulness of choosing the asymptotic over the exact form is due to convenience as it separates out the dependencies on $s$ (Rossby wave degree), $\ell^\prime-\ell$ (difference between harmonic degrees of the acoustic modes of interest) and radius into a product of a function $f_{\ell^\prime-\ell,s}$  and the kernel $\mathcal{K}_{n\ell}(r)$. We demonstrate the validity of this assumption in Figure \ref{fig:kernelAsymp} \citep[also see][]{hanasoge17_etal}. We choose two different cases, $\ell^\prime-\ell=1$ and $\ell^\prime-\ell=3$ and compare kernels obtained using the asymptotic expression with the corresponding exact forms. It is seen that these asymptotic kernel is a valid choice for our problem as the maximum harmonic degree of Rossby modes $s$ ($<20$) that we are considering is less than the minimum harmonic degree of the acoustic modes ($\ell=50$) used here. 
 
 In our previous work \citep{hanasoge19,mandal2020}, we were unable to infer even-degree Rossby modes as we only considered coupling between same-degree p-modes (see condition in Equation \ref{eq:fS}). Couplings between p modes with $\Delta\ell = \ell^\prime-\ell=1$ and $3$ enable us to detect Rossby waves of even harmonic degrees. In order to estimate $w^\sigma_{st}$ from the measured B-coefficients, $B^\sigma_{st}$, we need to perform inversions. We use the Regularized Least Squares (RLS) method discussed in \citet{mandal2020} to invert Equation \ref{eq:b_Coef}.
 
 \section{Results}\label{sec:results}
\begin{figure*}
    \centering
    \includegraphics[scale=0.45]{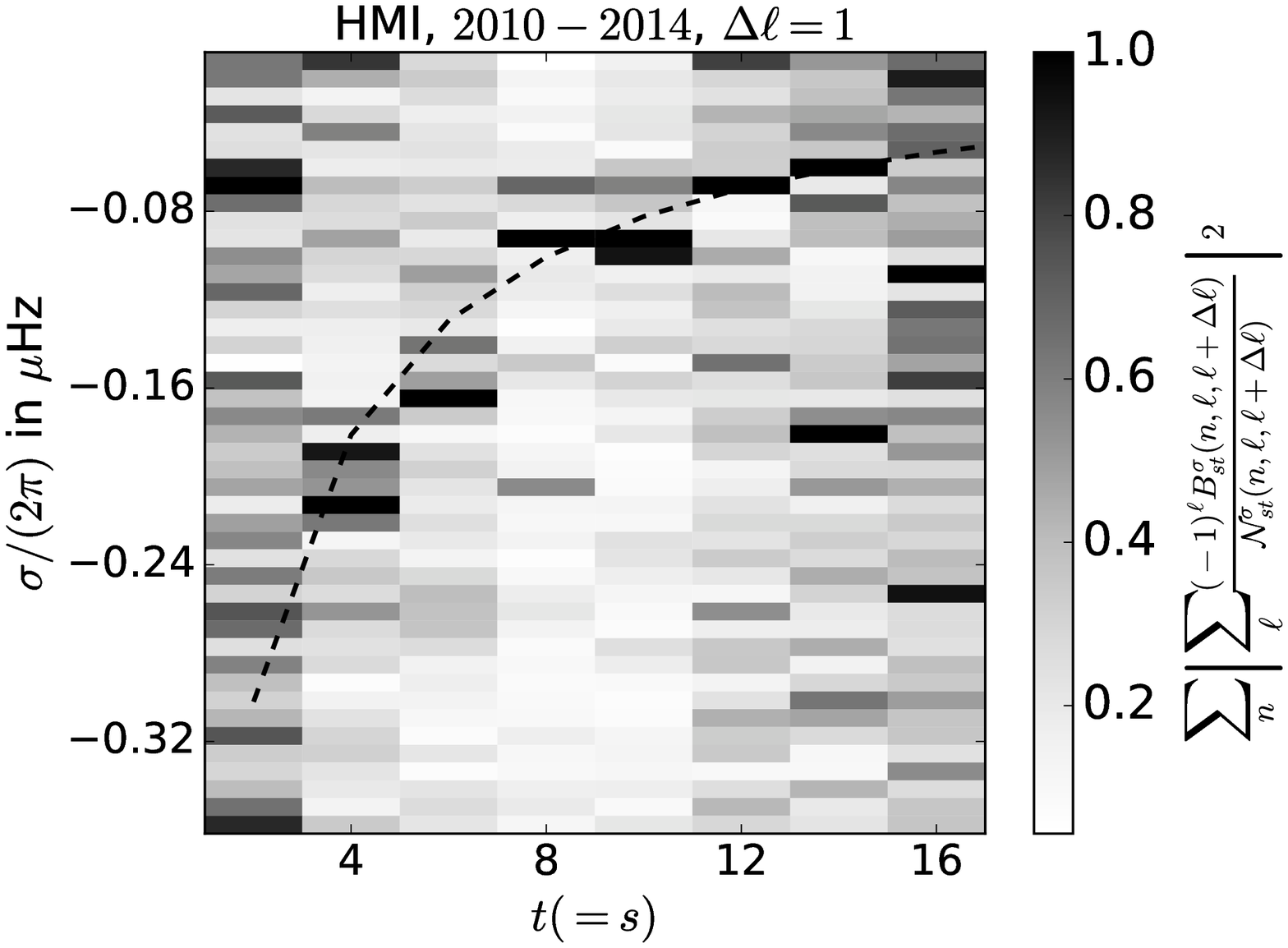}\includegraphics[scale=0.45]{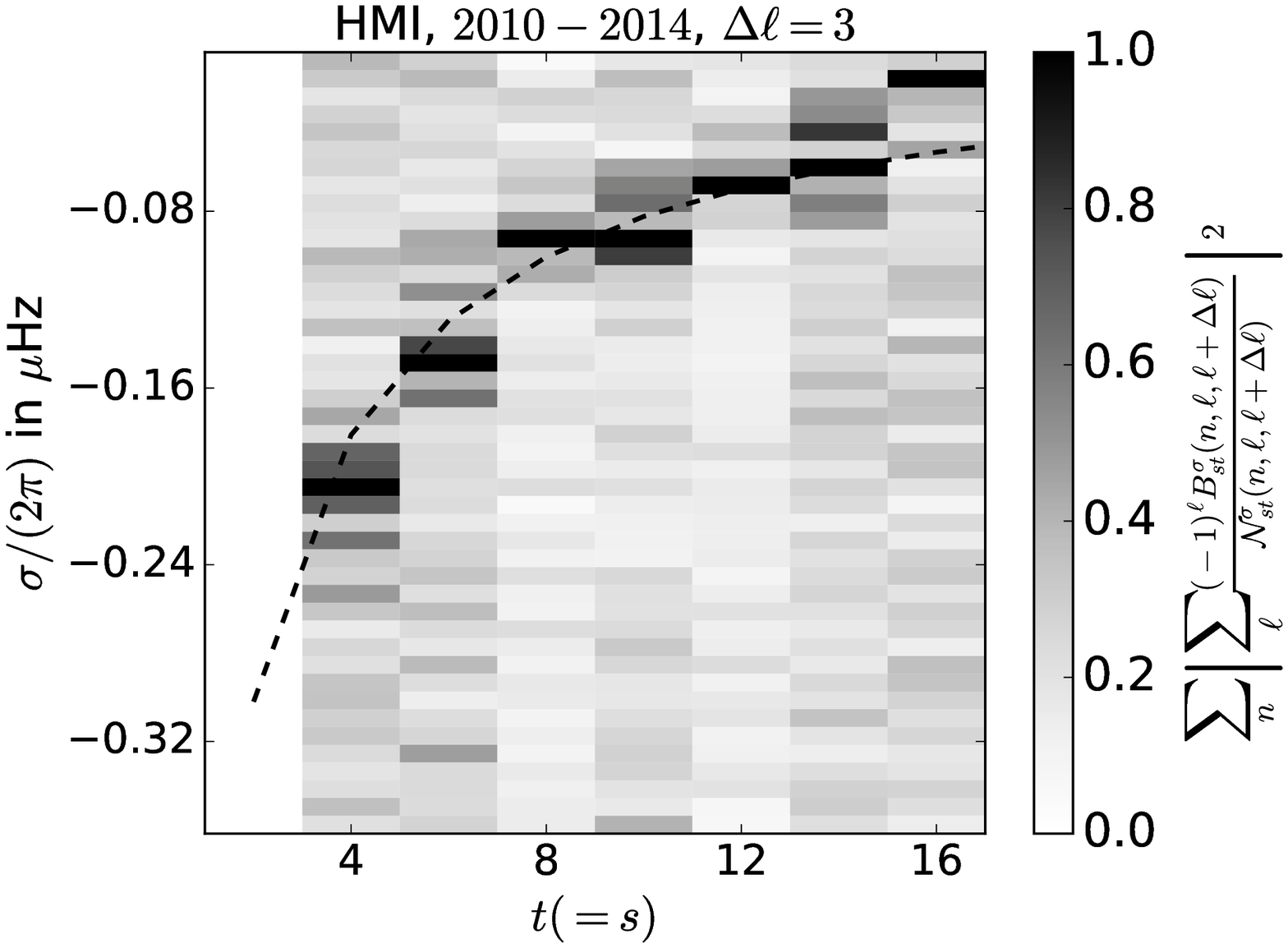}
    \includegraphics[scale=0.45]{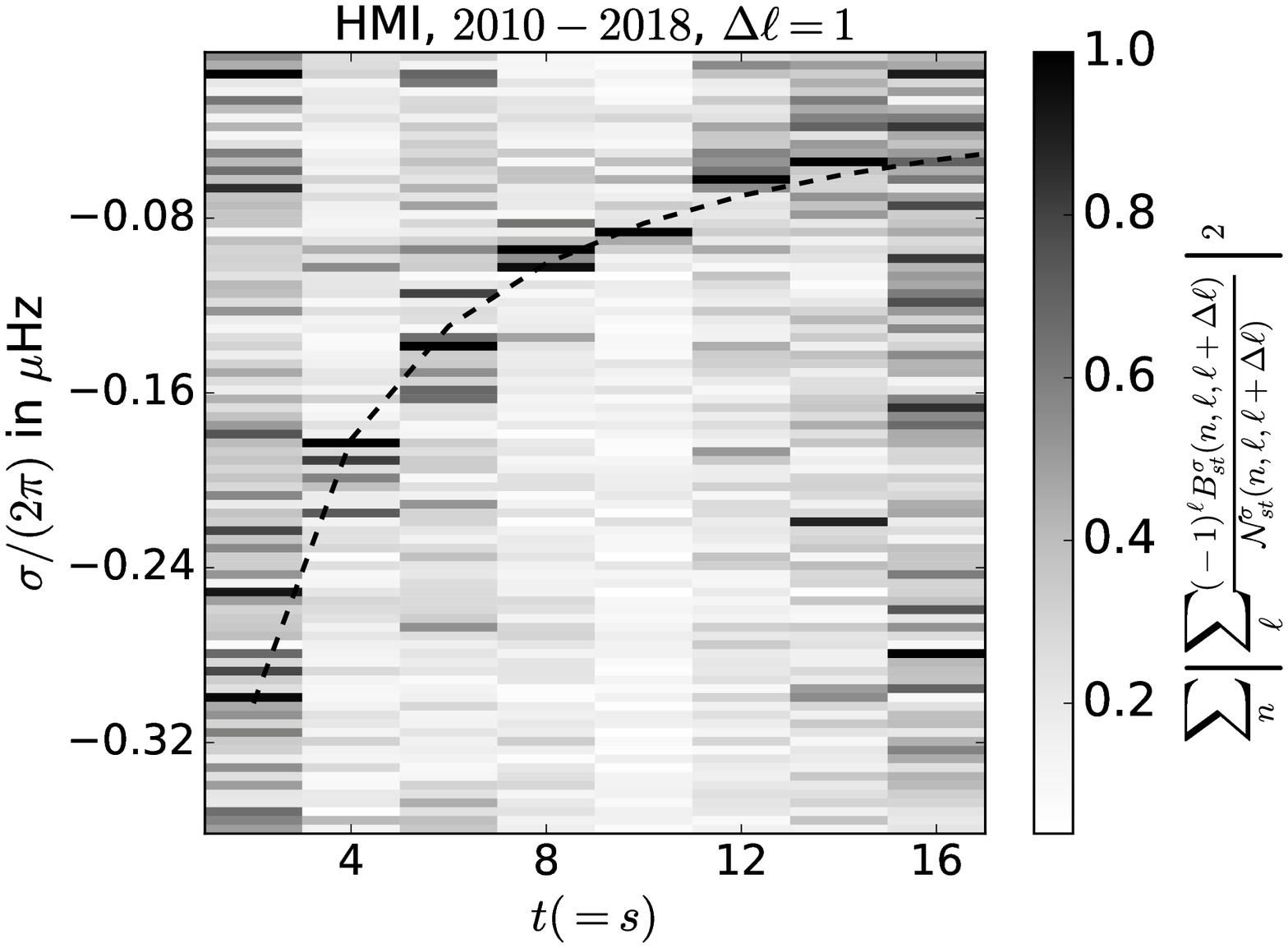}\includegraphics[scale=0.45]{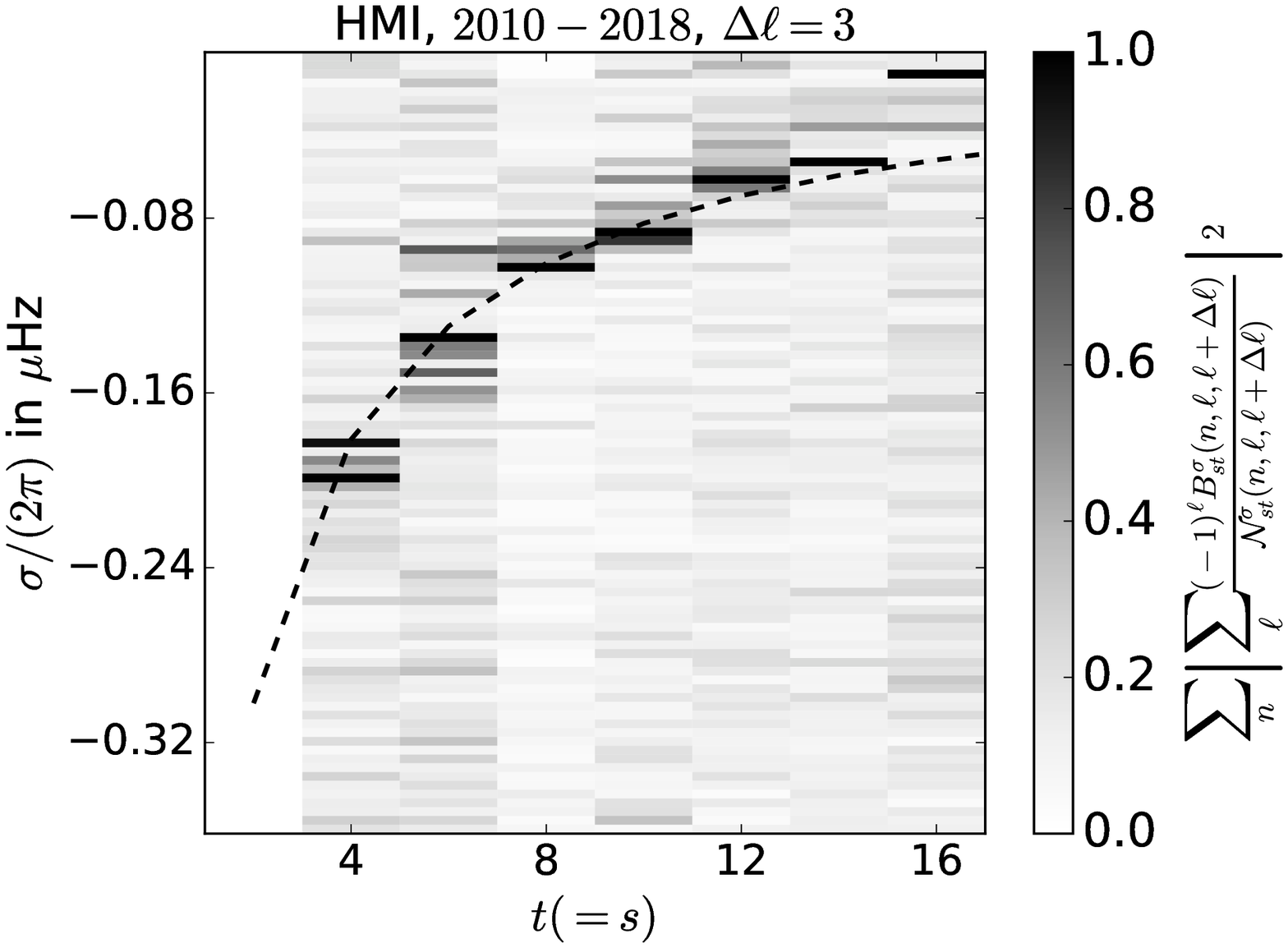}
    \caption{We estimate the weighted sum $\sum_{n}\vert\sum_{\ell} {(-1)^\ell B^\sigma_{st}(n,\ell,\ell+\Delta\ell)}/{\noise^{\sigma}_{st}(n,\ell,\ell+\Delta\ell)}\vert^2$ using the observed B-coefficients obtained after considering the coupling between acoustic modes with $\Delta\ell=1$ (left panel) and $3$ (right). Upper and lower panels show results from analyses of $4$ and $8$ years of SDO/HMI data, respectively.
    Power for each azimuthal order, $t$, has been normalized. The black dashed line is the theoretical dispersion relation for a uniformly rotating fluid with rotation frequency $\Omega/(2\pi)=453$ nHz. Modes with even azimuthal orders ranging from $t=2$ to $16$ are shown here.}
    \label{fig:bcoefPower}
\end{figure*}

  The Rossby signal in the measured B-coefficients may be calculated by performing a weighted sum over the signed B-coefficients $(-1)^\ell B^\sigma_{st}(n,\ell,\ell+\Delta\ell)$, over all harmonic degrees, $\ell$. We then take the sum of the squares of their absolute values over all radial orders, $n$, to estimate the quantity $\sum_{n}\vert\sum_{\ell} (-1)^\ell B^\sigma_{st}(n,\ell,\ell+\Delta\ell)/\noise^{\sigma}_{st}(n,\ell,\ell+\Delta\ell)\vert^2$, where $\noise^{\sigma}_{st}(n,\ell,\ell+\Delta\ell)$ corresponds to measurement noise. If the measured B-coefficients capture the signal properly, we would expect to see significant power close to the theoretical dispersion relation (Equation \ref{eq:dispersion2}), which is indeed the case, as seen in Figure \ref{fig:bcoefPower}. We show results for $\Delta\ell=1$ and $\Delta\ell=3$ separately. The sectoral mode $s=2$ does not appear when considering of $\Delta\ell=3$ since finite couplings are only possible when $s\geq \Delta\ell$. Therefore it can only be captured from measurements with $\Delta\ell=1$.\par     
\begin{figure*}
    \centering
    \includegraphics[scale=0.45]{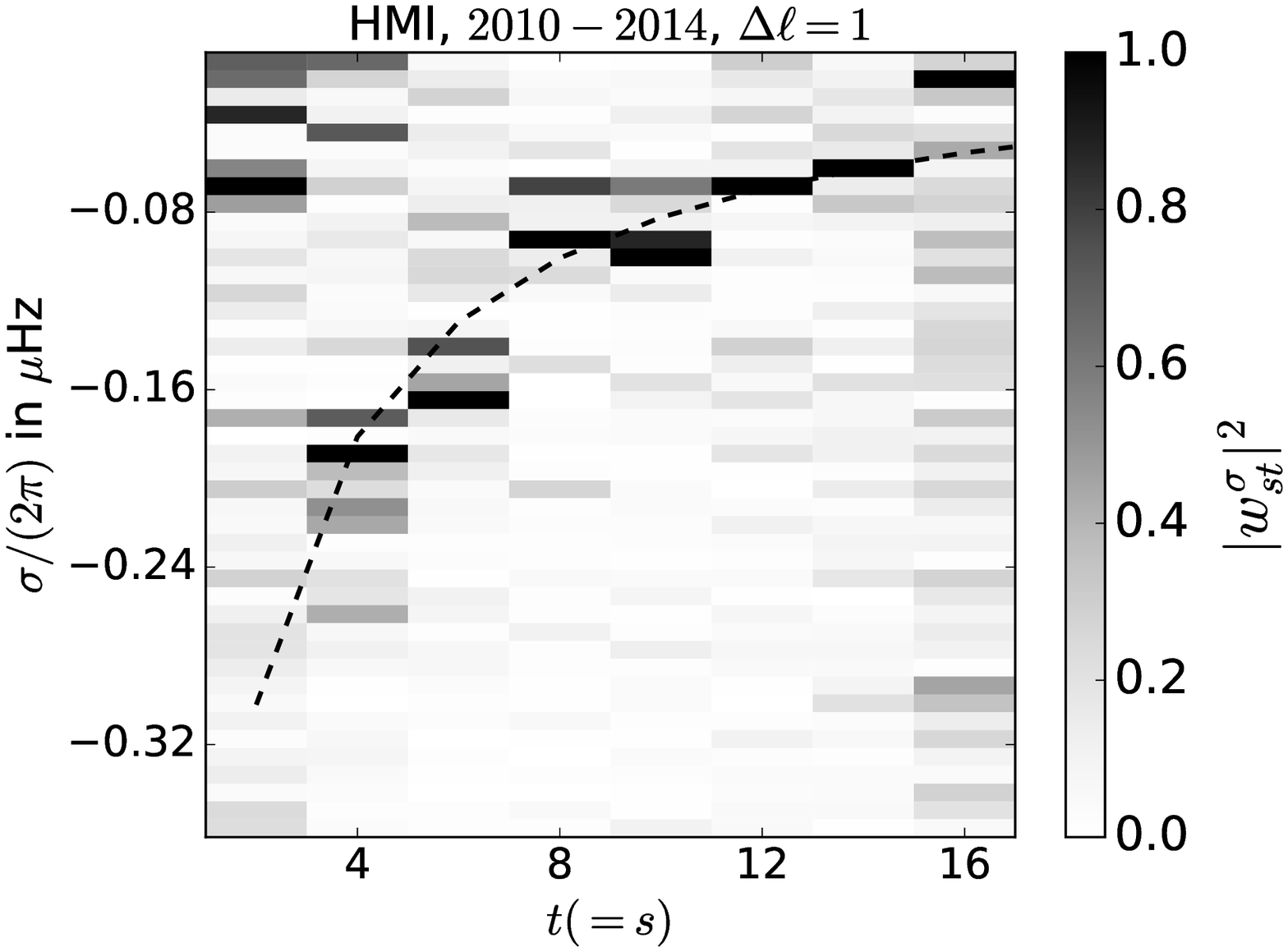}\includegraphics[scale=0.45]{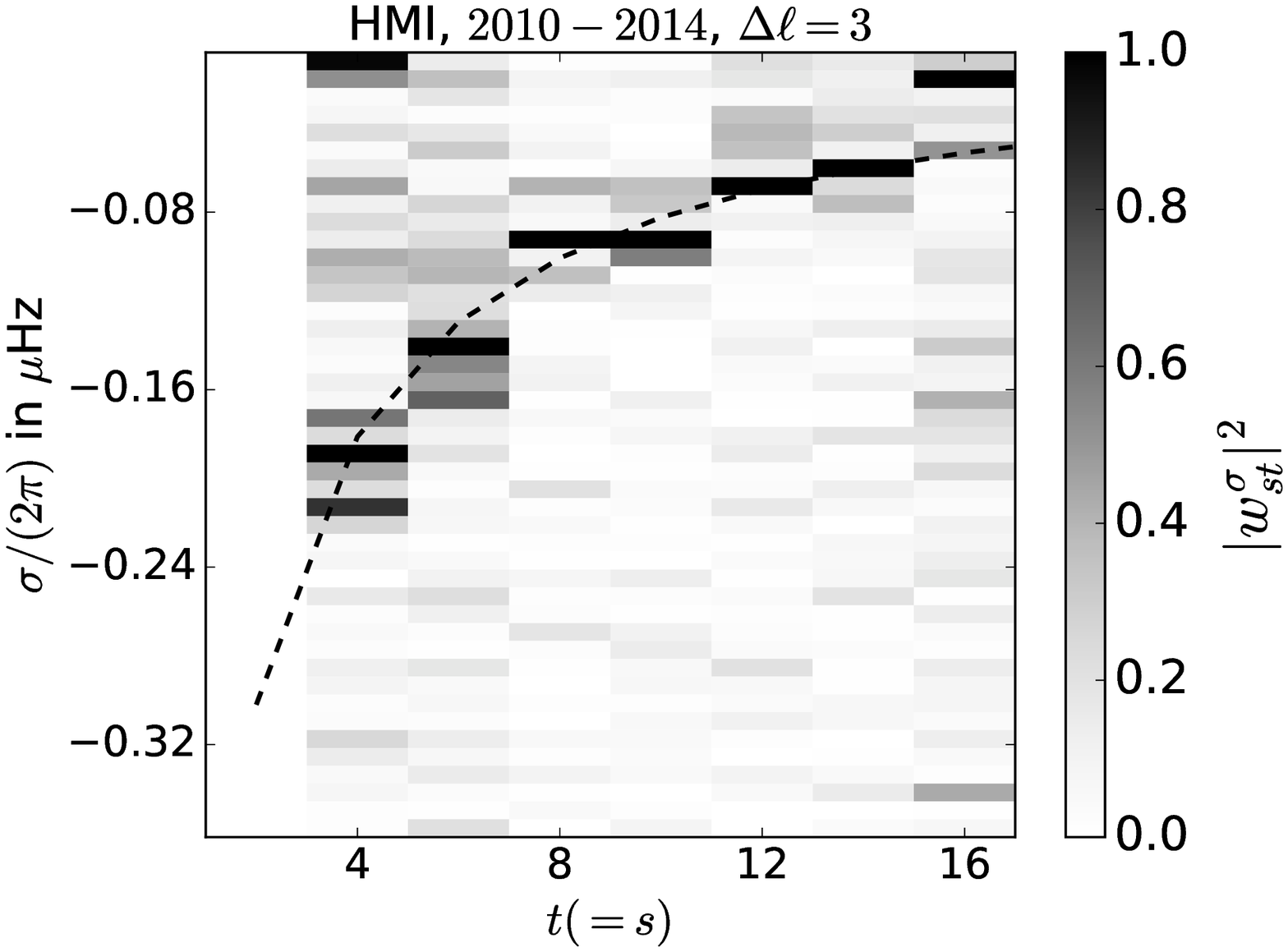}
    \includegraphics[scale=0.45]{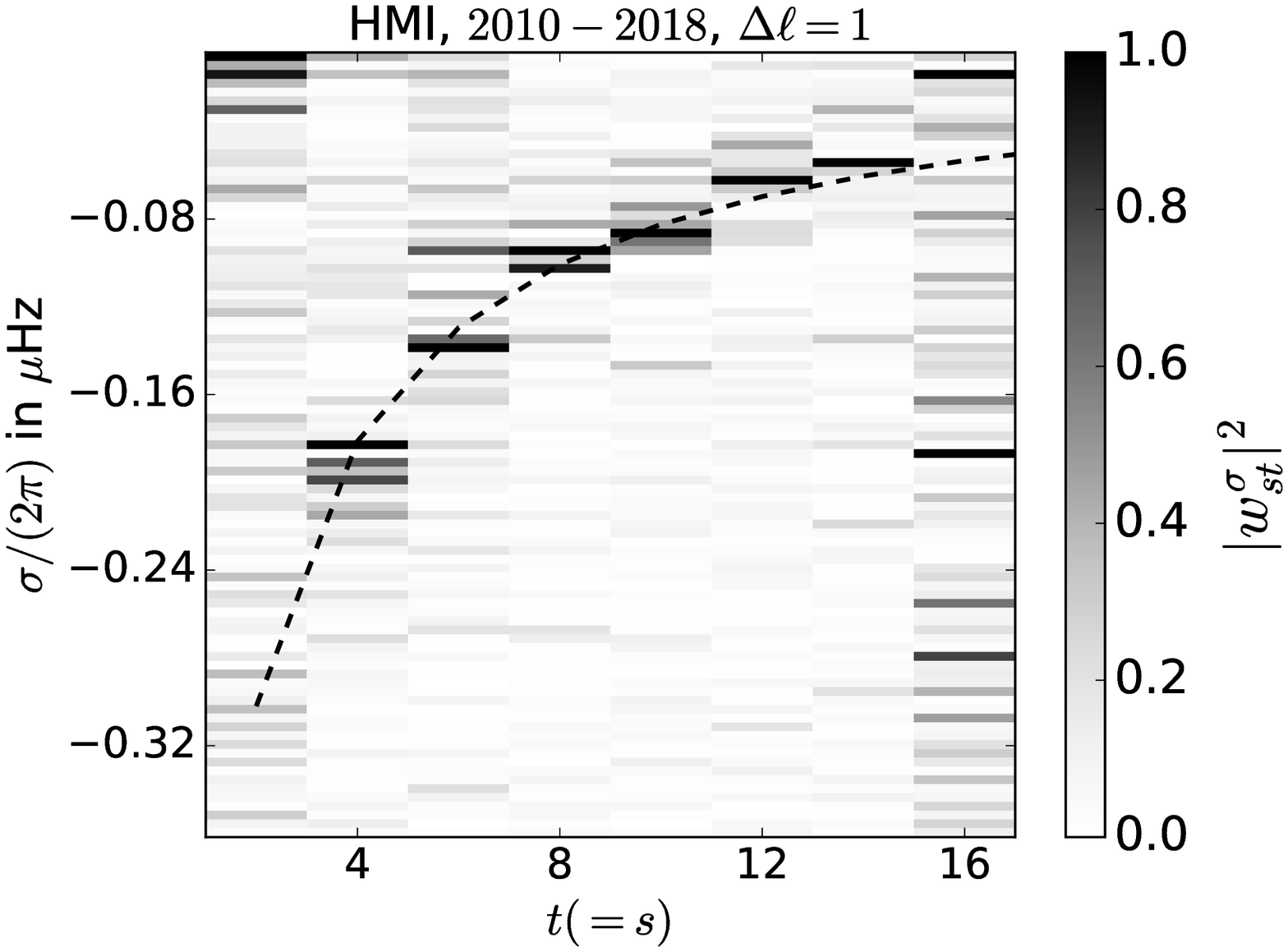}\includegraphics[scale=0.45]{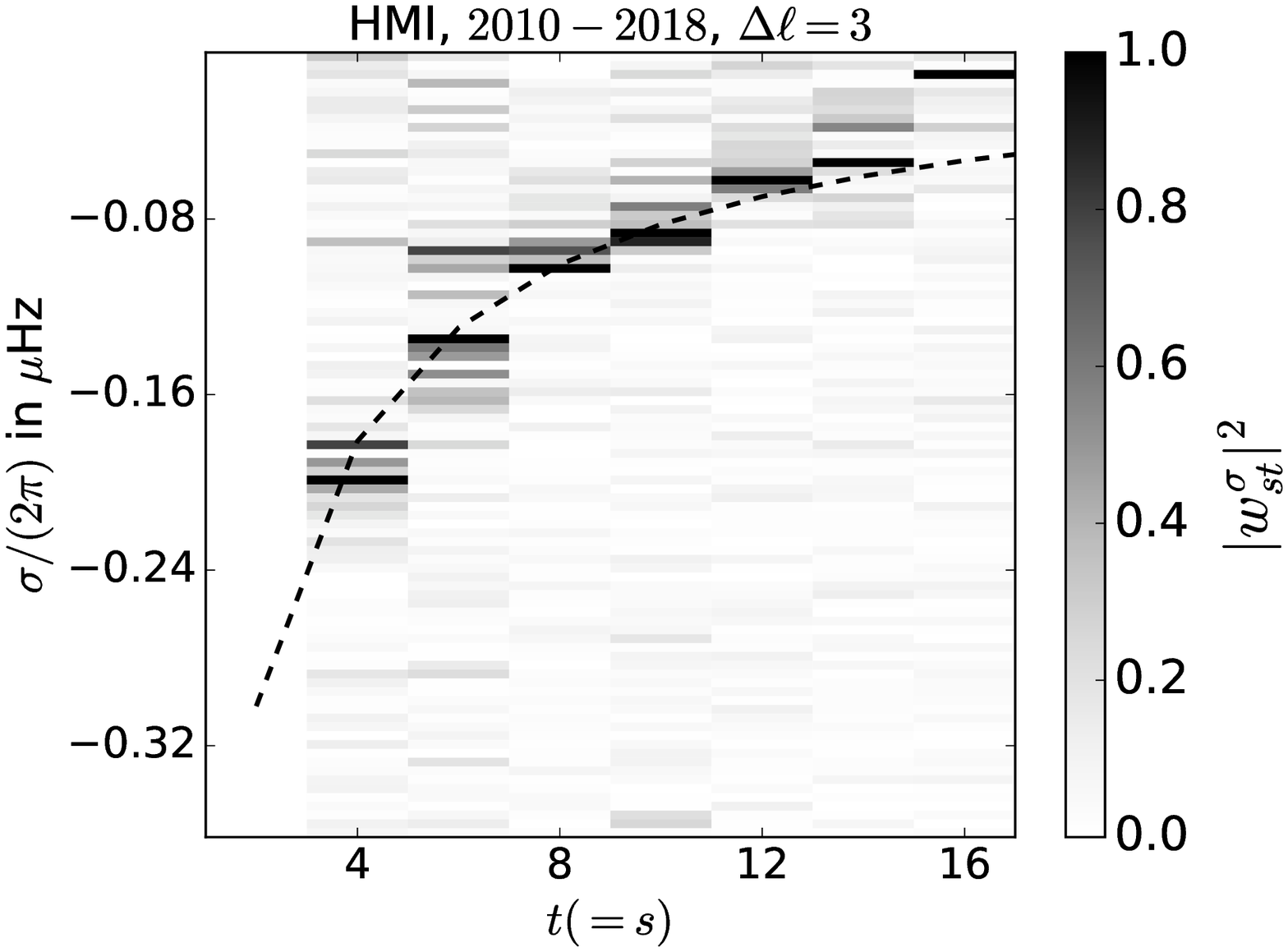}
    \caption{Normalized power, $\vert w^{\sigma}_{st}\vert^2$ of sectoral Rossby modes with azimuthal orders $t=2$ to $16$ obtained from inverting B-coefficients at depth $0.98R_\odot$. The black dashed line is the theoretical dispersion relation of Rossby modes (cf. Equation \ref{eq:dispersion2}) for a uniformly rotating medium at rate $\Omega/(2\pi)=453$ nHz. Upper and lower panels show results from analyses of $4$ and $8$ years of SDO/HMI data respectively, also mentioned in the titles of each panel. Left and right panels display results for $\Delta\ell=1$ and $\Delta\ell=3$ case respectively.}
    \label{fig:wstPower}
\end{figure*}

\begin{figure}
    \centering
    \includegraphics[scale=0.45]{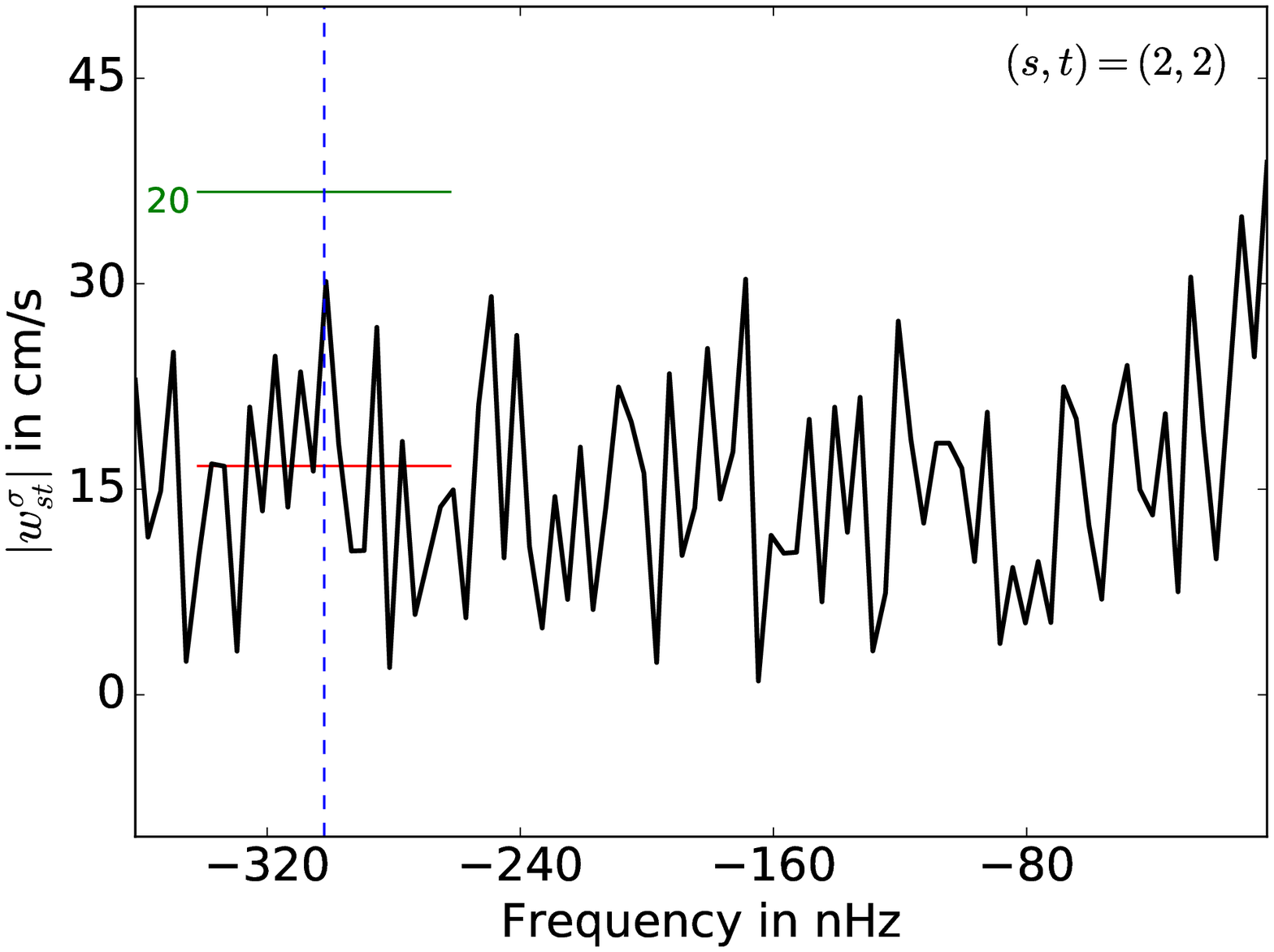}
    \caption{Power spectrum for Rossby mode $(s,t)=(2,2)$ from analyses of $8$ years of SDO/HMI data. The blue-dashed vertical line indicates the theoretically anticipated frequency of the $s=2$ sectoral mode. The red-solid line indicates the background power over the frequency range $[-342,-262]$ nHz. The green line corresponds to the power of the sectoral Rossby mode $s=2$ with amplitude $20$ cm/s.}
    \label{fig:s_2}
\end{figure}

   We obtain $w^\sigma_{st}$ (Equation \ref{eq:flow}) after performing inversions described in Section \ref{sec:inversion} and plot $|w^{\sigma}_{st}|^2$ at the depth $r=0.98R_\odot$ in Figure \ref{fig:wstPower}. The spectra show significant power close to the theoretical dispersion relation, similar to Figure \ref{fig:bcoefPower}, for all even azimuthal orders  $t=4$ to $16$. We do not detect power in the $t=2$ sectoral mode, as seen in Figure \ref{fig:s_2}. Power close to the theoretical frequency of this mode does not stand out from the background. Prior works by \citet{gizon18}, \citet{liang_2018}, \citet{proxauf2020} and \citet{hanson20} have also reached a similar conclusion. In order to place constraints on the amplitude of this mode, we highlight the anticipated $t=2$ Rossby-mode frequency in Figure \ref{fig:s_2}, where no peaks rise beyond the background. Based on the strength of the background power, the non-detection of this mode implies an upper amplitude limit of $0.2$ m/s. We also see significant power for sectoral mode $t=16$, though it is slightly shifted from the theoretical frequency. Indeed, we do not expect all the modes to exactly follow the analytical dispersion relation (Equation \ref{eq:dispersion2}) since it is derived for the case of a uniformly rotating sphere, whereas the Sun shows radial and latitudinal differential rotation. Other factors, e.g., convection and magnetic field, are not accounted for in Equation \ref{eq:dispersion2}. The detection of the $s=16$ mode, were it to be independently confirmed, would highlight the resolving power of this technique, since previous studies have not observed it.
  
\begin{figure*}
    \centering
    \includegraphics[scale=0.5]{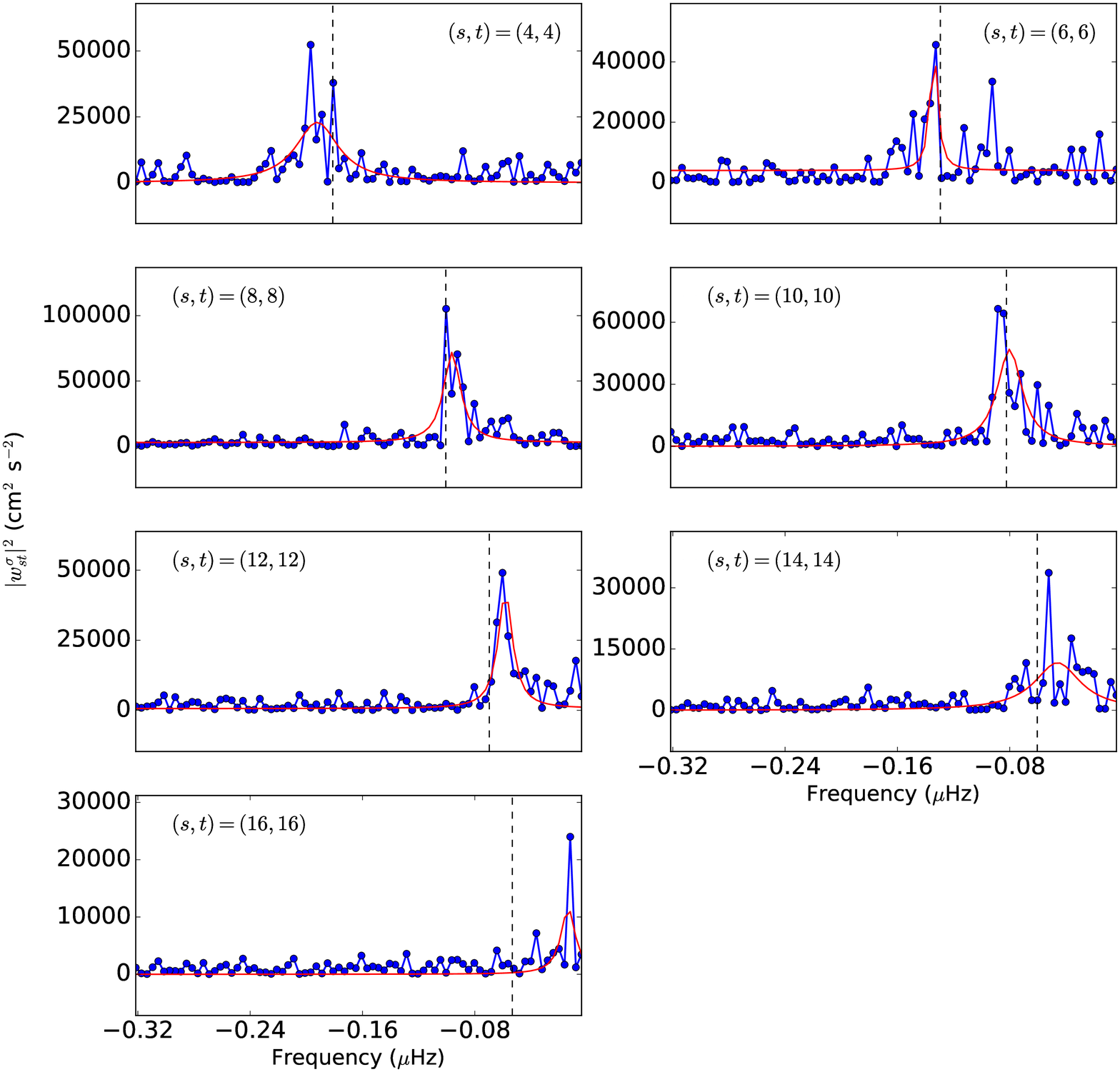}
    \caption{Analyses from $8$ years of SDO/HMI data. Power spectra of even-degree sectoral Rossby modes (value mentioned in each panel) at depth $0.98R_\odot$ are shown by blue solid lines with circles. Lorentzian function with a constant background, as described in Section \ref{sec:results}, have been fitted to these spectra to obtain mode frequency, line-width and amplitude. These fits are tabulated in Table \ref{tab:freq_HMI}. The fit spectrum is shown by the red solid line.}
    \label{fig:cutPower}
\end{figure*}

 After obtaining the profile of $w^\sigma_{st}$ from inversions, we fit a Lorentzian function with a constant background,
 \begin{equation}
     F(\sigma)=\frac{A}{1+[(\sigma-\sigma_0)/(\Gamma/2)]^2}+D,
 \end{equation}
 to $\vert w^\sigma_{s\,t} \vert^2$ of these sectoral modes for each harmonic degree, $s$. Here $A$, $\sigma_0$ and $\Gamma$ are amplitude, frequency and line-width of the mode respectively. $D$ is the constant background power. We use power spectra obtained from $8$ years of HMI data and the $\Delta\ell=3$ case for fitting. We follow the approach of \citet{anderson_1990} and minimize the following function 
 \begin{equation}
     \sum_{\sigma} \ln{F(\sigma)}+\frac{\vert w^\sigma_{s\,t}\vert^2}{F(\sigma)},\label{eq:fitFunc}
 \end{equation}
 with respect to the model parameters, $A$, $\sigma_0$, $\Gamma$ and $D$. We apply the {\it {fmin}} subroutine, which uses the downhill simplex algorithm, implemented in the {\it {scipy.optimize}} function to minimize Equation \ref{eq:fitFunc}. Values of these mode parameters obtained from the fitting are tabulated in Table \ref{tab:freq_HMI}. Fit spectra are shown in Figure \ref{fig:cutPower}. We compare the mode frequencies from our work with prior results of \citet{gizon18,liang_2018} in Table \ref{tab:freq_HMI}. We find that the Rossby-mode power peaks around $t=8$, which is consistent with the previous studies by \citet{gizon18}  and \citet{liang_2018}.
{
\renewcommand{\arraystretch}{1.5}
\begin{table*}
\caption{Measured values of Rossby mode parameters.}
\begin{center}
\begin{tabular}{c c c c c c c}
\hline
& &MC& \citet{gizon18} & \citet{liang_2018}\\
(s,t)&$-\frac{2\Omega /2\pi}{s+1}$&$\sigma_0/2\pi$ &$\sigma_0/2\pi$ &$\sigma_0/2\pi$&$\Gamma/2\pi$& $\sqrt{A}$ \\
 &nHz&nHz&nHz&nHz&nHz&cm s$^{-1}$\\
\hline
(4,\,4)&$-181$&$-192\,\pm\,4$&$-194^{+5}_{-4}$&$-198\,\pm\,5$&$28\,\pm\,4$&$151\,\pm\,68$\\
(6,\,6)&$-129$& $-134\,\pm\,2$&$-129\,\pm\,8$&$-135^{+4}_{-5}$&$5\,\pm\,1$&$205\,\pm\,70$\\
(8,\,8)&$-100$&$-96\,\pm\,3$&$-90\,\pm\,3$&$-91\,\pm\,3$&$11\,\pm\,5$&$262\,\pm\,69$\\
(10,\,10)&$-82.4$&$-80 \,\pm\,3$ &$-75\,\pm\,5$&$-60^{+5}_{-6}$&$21\,\pm\,1$&$216\,\pm\,89$\\
(12,\,12)&$-69.6$&$-58\,\pm\,4$&$-59\,\pm\,6$&$-36\,\pm\,8$&$12\,\pm\,1$&$204\,\pm\, 47$\\
(14,\,14)&$-60.4$&$-46\,\pm\,5$&$-56^{+6}_{-7}$&$-35\,\pm\,5$&$39\,\pm\,8$&$107\,\pm\,57$\\
(16,\,16)&$-53.2$&$-14\,\pm\,2$&—&—&$12\,\pm\,1$&$109\,\pm\,40$\\
\hline
\end{tabular} 
\tablefoot{ Analysis of first $8$ years of SDO/HMI data. Fitted values of Mode frequency, $\sigma_0$, amplitude $\sqrt{A}$, full width at half maximum $\Gamma$ to the observed $B$-coefficient spectra are listed in the co-rotating frame. MC stands for mode coupling here. The theoretical frequencies of modes in a co-rotating frame with tracking frequency $453$ nHz are given in the second column in the table.  Observed frequencies from two other studies, \citet{gizon18} and \citet{liang_2018} are listed in the fourth and fifth column in the table respectively. We plot the fitted spectrum in Figure~\ref{fig:cutPower}.}
\label{tab:freq_HMI}
\end{center}
\end{table*}
}

\section{Discussion and conclusions}
  Global-mode coupling, although a well known method, is still relatively unexplored in its applications to helioseismology.  Given the challenges in understanding the data and given that it is a relatively novel measurement process, our prior work focused on same-degree couplings, which have the additional benefit of mitigating the leakage of background power into the resonances. However, this limited us to being able to only draw inferences of odd-harmonic-degree toroidal flows. We expand the set of inferences here by considering coupling between p modes with finite harmonic-degree separation, $\Delta\ell = 1, 3$. This allows us to detect sectoral Rossby modes with even azimuthal numbers in the Sun over the range $t=4$ to $16$. Similar to earlier findings \citep{gizon18,liang_2018,proxauf2020,hanson20}, we also do not find evidence for the sectoral mode, $t=2$. There is no current understanding why the $t=2$ mode is undetectable. It is evident that power in this particular mode is very small or simply does not exist (see Figure \ref{fig:s_2}), which is of  interest in the context of understanding Rossby-mode excitation and dissipation. We also see evidence of sectoral-mode power beyond azimuthal number $t=15$. Though we consider two cases $\Delta\ell=1$ and $3$ in this work, other cases, e.g., $\Delta\ell>3$ may also be considered for this study, keeping in mind one caveat: modes with harmonic degree $s<\Delta\ell$ will not be captured due to the selection rule discussed in Section \ref{sec:data}. 
  We find that signal to noise ratio is better for  $\Delta\ell=3$ than $\Delta\ell=1$ (see Figure \ref{fig:wstPower}). We will investigate how signal to noise ratio varies for different $\Delta\ell$ in future work and whether it can be improved by combining all the seismic data together in the analysis instead of separately, as done here. \par
 We present our results only for near-surface layers in this work. However, because we analyze couplings between global modes, we are able to infer $w^\sigma_{st}$ as a function of radius after inversion. The inference of the radial dependence of these modes at both odd and even azimuthal numbers is deferred to future work; this may shed light on the excitation mechanism of these modes. Although \citet{mandal2020} have shown that spatial leakage does not affect inferences of the radial dependencies of these modes, care must still be taken as the observed data might contain other systematical errors, which may influence the inferences. This work, taken together with our previous work \citep{hanasoge19}, highlights mode-coupling as very useful in determining length and time scales of multi-scale dynamics in the Sun. \par

 \textit{Acknowledgments:} We thank the referee for useful comments which helped to improve the manuscript. K.M. and L.G. acknowledge support from ERC Synergy grant WHOLESUN  810218. S.M.H. acknowledges the Max Planck Partner Group program.
%
\bibliographystyle{aa}
 \bibliography{referencesKri}
\end{document}